# Pressure-induced reemergence of superconductivity in topological kagome metal $CsV_3Sb_5$

Zhuyi Zhang,[1,2,§] Zheng Chen,[1,2,§] Ying Zhou,[3] Yifang Yuan,[1,2] Shuyang Wang,[1,2] Jing Wang,[1,2] Haiyang Yang,[1,2] Chao An,[3] Lili Zhang,[4] Xiangde Zhu,[1] Yonghui Zhou,[1] Xuliang Chen,[1,*] Jianhui Zhou,[1,†] and Zhaorong Yang[1,3,5,‡]

[1]*Anhui Key Laboratory of Condensed Matter Physics at Extreme Conditions, High Magnetic Field Laboratory, HFIPS, Anhui, Chinese Academy of Sciences, Hefei 230031, P. R. China*

[2]*Science Island Branch of Graduate School, University of Science and Technology of China, Hefei 230026, China*

[3]*Institutes of Physical Science and Information Technology, Anhui University, Hefei 230601, China*

[4]*Shanghai Synchrotron Radiation Facility, Shanghai Advanced Research Institute, Chinese Academy of Sciences, Shanghai 201204, China*

[5]*Collaborative Innovation Center of Advanced Microstructures, Nanjing University, Nanjing 210093, China*

§These authors contribute equally to this work.
*Corresponding author: xlchen@hmfl.ac.cn;
†Corresponding author: jhzhou@hmfl.ac.cn;
‡Corresponding author: zryang@issp.ac.cn.

Quasi-two-dimensional kagome metals $AV_3Sb_5$ (A = K, Rb, and Cs) have attracted much recent interest due to exotic quantum phenomena such as unconventional superconductivity, topological charge order and giant anomalous Hall effect. Here we report pressure-induced reemergent superconductivity in $CsV_3Sb_5$ by electrical transport measurements under high pressures up to 47.9 GPa. We show that the superconducting critical temperature $T_c$ is first enhanced by pressure and reaches its first maximum ~ 8.9 K at 0.8 GPa, then the $T_c$ is suppressed by pressure and cannot be detected above 7.5 GPa, forming a dome-shaped superconducting phase diagram. Remarkably, upon further compression above 16.5 GPa, a new superconducting state arises, of which $T_c$ is enhanced by pressure to a second maximum ~ 5.0 K and the reemergent superconductivity keeps robust up to 47.9 GPa. Combined with high-pressure synchrotron x-ray diffraction measurements that demonstrate the stability of the pristine hexagonal phase up to 43.1 GPa, we suggest that the reemergence of superconductivity in the V-based superconductor could be attributed to a pressure-induced Lifshitz transition.

# I. INTRODUCTION

Recently, a new class of quasi-two-dimensional topological Kagome metals $AV_3Sb_5$ (A= K, Rb, and Cs) has been attracting great interest. Combination of topologically nontrivial electronic structures of these compounds and strongly correlated effects leads to a series of fascinating quantum phenomena [1-15], such as superconductivity, charge density wave (CDW), and giant anomalous Hall effect. Scanning tunneling microscope/spectroscopy (STM/STS) revealed that the unconventional strong-coupling superconductivity and pair density wave coexist in $CsV_3Sb_5$ [1]. A robust zero-bias conductance peak inside the superconducting (SC) vortex core was observed in $CsV_3Sb_5$, implying topological superconductivity [2]. First-principles calculations suggest that the weak electron-phonon coupling in $AV_3Sb_5$ hardly accounts for the superconductivity, in agreement with its nature of unconventional superconductivity [3]. High-resolution STM also demonstrated the existence of topological CDW state in $KV_3Sb_5$ [4]. A remarkably giant anomalous Hall conductivity in $KV_3Sb_5$ and $CsV_3Sb_5$ could reach the magnitude of $10^4$ $\Omega^{-1}$cm$^{-1}$, which may come from the skew scattering [5,6], or the chiral flux phase [7]. The nontrivial electronic topology and giant anomalous Hall effect together with the correlated states (superconductivity and CDW) make $AV_3Sb_5$ a unique platform to investigate rich quantum states.

Pressure is an effective method to tune the lattice structure and to manipulate electronic state such as spin/charge density wave, magnetism, superconductivity, and topological phases of matter without introducing impurities. It has been shown that pressure can suppress the CDW and induce superconductivity [16]. In most cases, the SC critical temperature $T_c$ within a finite pressure range usually forms a single dome or decreases monotonically. Interestingly, multiple SC domes (e.g. two domes), a strong precursor of unconventional superconductivity, was reported in various materials [17], such as $A_{1-x}Fe_{2-y}Se_2$ (A= K, Rb, Tl/K, or Tl/Rb) [18], $(Li_{1-x}Fe_x)OHFe_{1-y}Se$ [19], $CeCu_2Si_2$ [20] and FeS [21]. Whether $AV_3Sb_5$ could support the unconventional multiple-dome superconductivity under pressure is highly desired.

In this work, we report the pressure-induced two-dome superconductivity in $CsV_3Sb_5$, a material whose SC critical temperature $T_c$ is 2.5 K at ambient pressure [11]. Under high pressures, $T_c$ is initially enhanced to 8.9 K at 0.8 GPa and then suppressed gradually, forming a dome (SC-I) in the *pressure-temperature* phase diagram. Another SC dome (SC-II) emerges above 16.5 GPa, of which the $T_c$ increases upon compression and later the reemergent superconductivity keeps robust up to 47.9 GPa. High-pressure synchrotron x-ray diffraction (XRD) reveals no crystal structure phase transition. This two-dome superconductivity in $CsV_3Sb_5$ can be attributed to a pressure-induced Lifshitz transition in line with the *c/a* anomaly at ~ 8 GPa.

# II. METHODS

$CsV_3Sb_5$ single crystals were grown via self-flux method. The chemicals were mixed with the molar ratio of Cs: V: Sb = 1: 3: 20 in a MgO crucible, and then sealed in a vacuumed quartz tube. The whole process was performed in a glovebox with argon as protection gas. The tube was heated slowly in a box furnace at a rate of 1 ˚C/min to 1000 ˚C, then kept at this temperature for 24 hours for a thorough reaction. After that, the furnace was slowly cooled at 2 ˚C/h down to 640 ˚C. The extra flux was removed by fast centrifuging.

A standard four-probe method was employed to perform the high-pressure electrical transport measurements in a temperature range of 1.8–300 K in a Be-Cu diamond-anvil cell (DAC) using sodium chloride (NaCl) powder as the transmitting medium. The electric current was applied within the *ab* plane. The crystalline phase of CsV$_3$Sb$_5$ was checked by the single crystalline x-ray diffraction (XRD, Cu K$_\alpha$, $\lambda$ = 1.54184 Å), which displays a (00*l*) orientation of the single crystal cleavage surface. The chemical composition value of Cs: V: Sb as ~1: 2.87: 5.00 is given by Energy-dispersive x-ray spectra (EDX) [Fig. 1(c)] with area and point-scanning modes. High-pressure angle-dispersive synchrotron x-ray diffraction (XRD) experiments were carried out at room temperature with CsV$_3$Sb$_5$ powder crushed from single crystals at pressures up to 43.1 GPa, at the beamline BL15U1 of the Shanghai Synchrotron Radiation Facility (SSRF) ($\lambda$ = 0.6199 Å), using Daphne 7373 as the transmitting medium. The DIOPTAS [22] program was used for image integrations and the XRD patterns were fitted by using the RIETICA program [23] with Le Bail method. The pressure values for all of the above experiments were determined by the ruby fluorescence method [24].

## III. RESULTS AND DISCUSSION

***Sample characterization--*** CsV$_3$Sb$_5$ crystallizes in a hexagonal structure *P*6/*mmm* as shown in Fig. 1(a). In this V-based kagome metal, the vanadium sublattice in the V-Sb layers forms a two-dimensional kagome network, and the alkali metal ions are inserted between the V-Sb layers [9,11]. Figure. 1(b) displays the XRD pattern of a CsV$_3$Sb$_5$ single crystal. It is clear that only (00*l*) diffraction peaks can be detected which indicates the as-grown single crystal having a (00*l*) preferred orientation. The *c*-axis lattice constant is 9.23±0.05 Å, consistent with the previous value of CsV$_3$Sb$_5$ [11]. From the EDX data in Fig. 1(c), the chemical composition Cs: V: Sb = 1: 2.87: 5.00 implies slight V-vacancies in the as-grown single crystals. Figure. 1(d) displays the temperature dependence of resistivity of the sample with the SC transition at onset $T_c$ about 5.0 K and offset $T_c$ about 3.0 K, and a CDW anomaly at $T_{CDW}$~ 92 K under ambient pressure. Note that the variations of SC $T_c$ (2.5-5.0 K) at ambient pressure could come from the different contents of vanadium vacancies.

***Pressure-induced superconductivity--*** At the starting pressure of 0.8 GPa, no anomalies associated with the CDW transition are observed in the temperature dependence of resistance as shown in Fig. 2, which indicates that the CDW order could be highly sensitive to the application of pressure. In addition, one can see a remarkable drop of resistance at 0.8 GPa which starts at ~8.9 K and reaches zero at ~3.7 K as shown in Fig. 2(b). This is a sharp increment of onset $T_c$ compared with the value at ambient pressure. Above 0.8 GPa, $T_c$ turns to decrease with increasing pressure, and the sample loses its zero resistance at ca. 2.9 GPa [Fig. 2(b)]. Then, $T_c$ shifts to lower temperatures continuously until it is undetectable at 7.5 GPa down to 1.8 K [Fig. 2(c)]. The pressure range in which superconductivity cannot be detected extends to 13.3 GPa. At 16.5 GPa, a new drop of resistance appears at 3.1 K [Fig. 2(e)]. This resistance drop becomes more prominent with further compression, and reaches zero resistance eventually at ca. 36.4 GPa [Fig. 2(f)]. It is clear that the new resistance drop corresponds to a new SC transition. Furthermore, we find that the onset temperature shifts to higher temperatures with increasing pressure initially [Figs. 2(d)-1(f)], reaches its maximum (~5.0 K) at ~24.4 GPa, and then persists up to 47.9 GPa, the maximal pressure investigated in this work.

In order to confirm the SC nature of the new resistance drop observed in CsV$_3$Sb$_5$, we applied magnetic fields perpendicular to the *ab* plane of the sample at 1.7 GPa and 36.4 GPa [Figs. 3(a) and 3(b)]. It can be seen that with the increase of the magnetic field, both drops of resistance shift to lower temperatures and become almost suppressed under the magnetic field of 2.0 and 2.5 T for the compressed sample at 1.7 GPa and 36.4 GPa, respectively. By defining $T_c$ values under magnetic fields with the resistance criterion of $R_{cri} = 90\%R_n$ ($R_n$ is the normal-state resistance), we can extract the field ($H$) dependence of $T_c$ for CsV$_3$Sb$_5$ at 1.7 GPa and 36.4 GPa [Figs. 3(a) and 2(b)] and plot the $H_{c2}(T_c)$ data in Fig. 3(c). The experimental data is fitted with Werthamer-Helfand-Hohenberg (WHH) model [25]. The estimated upper critical magnetic field $H_{c2}(T_c)$ values at zero temperature are 2.09 T at 1.7 GPa, and 2.58 T at 36.4 GPa.

*High-pressure XRD--* In order to investigate the nature of the reemergence of superconductivity in pressurized CsV$_3$Sb$_5$, we performed *in situ* high-pressure XRD measurements in two separate runs as shown in Fig. 4(a) and 4(b). In Run1 (0.2-7.6 GPa), considering the $T_c$ is enhanced sharply from 5.0 K at ambient pressure to 8.9 K at 0.8 GPa, pressure was finely tuned with a minimal step about 0.3 GPa. In Run2, pressure was extended to 43.1 GPa. No structure phase transition was observed under pressures in both runs, and the lattice constants *a* and *c* decrease monotonically upon compression as shown in Figs. 4(c)-3(d). By fitting the evolution of volume *V/Z* (the unit-cell volume per chemical formula) to the third-order Birch-Murnaghan formula [26]:

$$P = \frac{3}{2}B_0\left[(V_0/V)^{\frac{7}{3}} - (V_0/V)^{\frac{5}{3}}\right] \times \{1 + \frac{3}{4}(B_0' - 4)\left[(V/V_0)^{\frac{2}{3}} - 1\right]\}, \quad (1)$$

where $V_0$, $B_0$, and $B_0'$ represent the zero-pressure volume, bulk modulus -*V/(dV/dP)*, and first-order derivative of the bulk modulus at zero pressure, respectively, then we obtain $V_0 = 236.6$ Å$^3$, $B_0 = 38.1$ GPa, and $B_0' = 6.5$, see Fig. 4(e). There are no extraordinary anomalies in the *V/Z-P* curve in Fig. 4(e). Moreover, the value of *c/a* ratio also decreases monotonically, while exhibits a kink at around 8 GPa as shown in Fig. 5(b), which is right located in the middle of the pressure range where the SC state is undetectable in the high-pressure resistance experiments.

*P-T phase diagram--* As shown in Fig. 5(a), the *Pressure-Temperature (P-T)* phase diagram reveals two distinct SC regions: the first dome-shaped SC state (SC-I), and the pressure-induced reemergent SC state (SC-II). In the SC-I dome, $T_c$ is firstly enhanced to 8.9 K upon compression at 0.8 GPa and then suppressed continuously until it is not detectable above ~ 7.5 GPa. In the SC-II dome, superconductivity reemerges at about 16.5 GPa, and $T_c$ increases upon compression, reaches a maximum (5.0 K) at 24.4 GPa. Upon further compression, SC-II remains anomalously robust up to 47.9 GPa.

*Origin of superconductivity--* Reemergence of superconductivity has been reported in various materials that show unconventional superconductivity and have different interpretations [17]. Crystal structure phase transition is usually responsible for two-dome superconductivity, such as in Sr$_{0.065}$Bi$_2$Se$_3$ [27] and LaFeAsO$_{1-x}$F$_x$ (0≤x≤0.75) [28]. However, since our XRD results reveal no structure phase transition up to 43.1 GPa, this mechanism can thus be ruled out. In many other materials exhibiting two SC domes, the first SC dome is usually attributed to the competition or coexistence between superconductivity and some order parameters (e.g. CDW, SDW, antiferromagnetic order). In CsV$_3$Sb$_5$, the first SC dome can be attributed to the competition of CDW and superconductivity [29-31].

However, due to the limitation of the high-pressure experimental technique, the origin of the second SC dome is still an enigma. Though the second SC dome has been associated with valance fluctuations in $CeCu_2Si_2$ [20], magnetic (e.g. antiferromagnetic) phases in $LaFeAsO_{1-x}H_x$ ($0 \leq x \leq 0.53$) [32], Lifshitz transition (or Fermi surface reconstruction) in $(Li_{1-x}Fe_x)OHFe_{1-y}Se$ [19] and FeS [33], a direct evidence is still lacking. Since no magnetic orders are reported in the family of $AV_3Sb_5$ [34], antiferromagnetic or ferromagnetic fluctuations turn out to be irrelevant to the two-dome superconductivity in $CsV_3Sb_5$. In fact, we cannot exclude the CDW fluctuation in triggering the second SC dome. Nevertheless, we notice that the $c/a$ ratio displays a kink anomaly at ~ 8 GPa, which is in line with the scenario of Lifshitz transition. The Lifshitz transition occurs when an extreme of the electronic band structure crosses the Fermi level, which is associated with a van Hove singularity in the density of states [35]. The Lifshitz transition does not require a structural transition but could lead to anomalies in the phonon spectrum, or in the $c/a$ ratio [36-40]. Accordingly, the second SC dome in $CsV_3Sb_5$ can thus be attributed to the Fermi surface reconstruction caused by the Lifshitz transition. In line with this scenario, a recent study revealed that at high pressure, the evolution of the Fermi surface governed by the disappearance and reappearance of Sb $p_z$ pockets could correlate with the suppression and re-entrance of SC in $CsV_3Sb_5$ [41].

## IV. CONCLUSIONS

By combining the high-pressure electronic transport and XRD, we found two-dome superconductivity in the topological kagome metal $CsV_3Sb_5$. The XRD results reveal no structure phase transition up to 43.1 GPa. The first SC dome comes from the competition between CDW and superconductivity. Meanwhile, the second SC dome is attributed to the Fermi surface reconstruction caused by the Lifshitz transition. Our work calls for a thorough theoretical study to uncover the nature of the reemergence of superconductivity under high pressure.

## ACKNOWLEDGMENTS

The authors are grateful for the financial support from the National Key Research and Development Program of China (Grants No. 2018YFA0305704), the National Natural Science Foundation of China (Grants No. 11804344, No. U19A2093, No. U1632275, No. U1932152, No. 11874362, No. 11704387, No. 11804341, and No. U1832209), the Natural Science Foundation of Anhui Province (Grants No. 1908085QA18 and No. 2008085QA40), the Users with Excellence Program of Hefei Center CAS (Grants No. 2020HSC-UE015 and No. 2021HSC-UE008), and the Collaborative Innovation Program of Hefei Science Center CAS (Grant No. 2020HSC-CIP014). A portion of this work was supported by the High Magnetic Field Laboratory of Anhui Province under Contract No. AHHM-FX-2020-02. Yonghui Zhou was supported by the Youth Innovation Promotion Association CAS (Grant No. 2020443). The x-ray diffraction experiment was performed at the beamline BL15U1, SSRF.

# Figures and captions

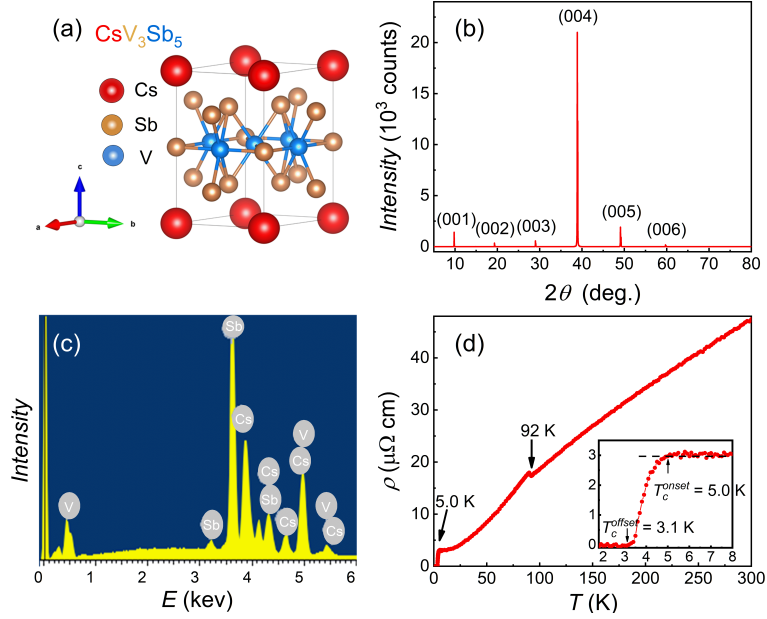

FIG. 1. (a) Schematic crystal structure of CsV$_3$Sb$_5$, with a space group of $P6/mmm$, where red, tan and blue spheres represent Cs, Sb, and V, respectively. (b) Single crystalline x-ray diffraction pattern with the corresponding Miller indices (00$l$). (c) Energy-dispersive x-ray spectroscopy of CsV$_3$Sb$_5$ single crystal. (d) Temperature dependence of resistivity at ambient pressure, in which the kink at ~ 92 K corresponds to the CDW order. The inset on the lower right indicates the SC transition at onset $T_c$ ~ 5.0 K and offset $T_c$ ~ 3.1 K.

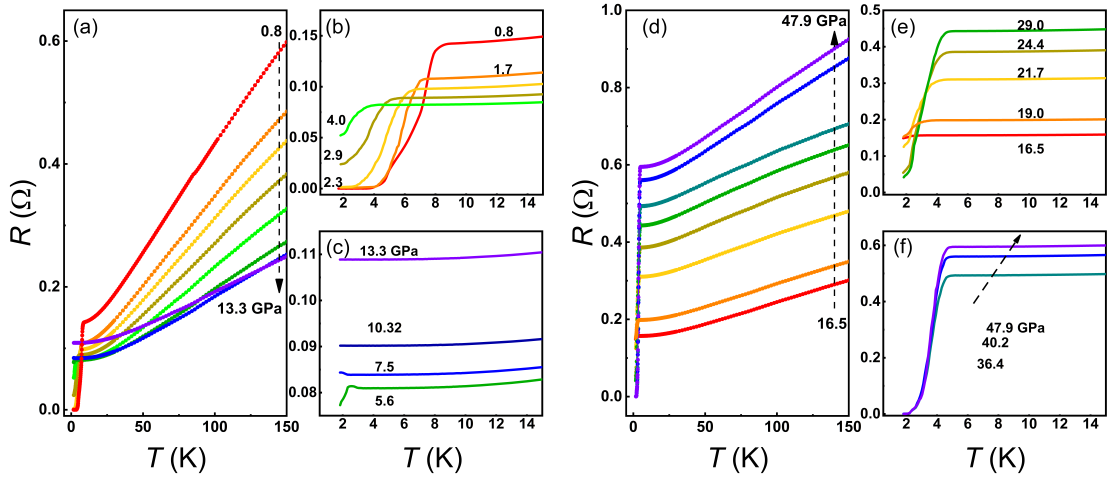

FIG. 2. Temperature-dependent in-plane resistance of CsV$_3$Sb$_5$ at various pressures. $R$-$T$ curves at temperature range of 1.8-150 K for (a) 0.8-13.3 GPa, and for (d) 16.5-47.9 GPa. Zoom-in of the low-temperature region for (b) 0.8-4.0 GPa, (c) 5.6-13.3 GPa, (e) 16.5-29.0 GPa, (f) 36.4-47.9 GPa.

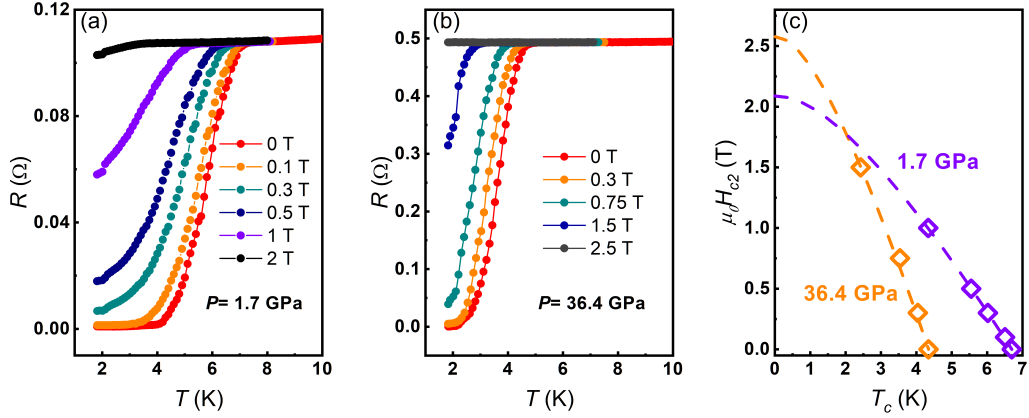

FIG. 3. Upper critical field in $CsV_3Sb_5$ under pressure. (a), (b) Temperature dependence of resistance under different magnetic fields perpendicular to the *ab* plane at 1.7 GPa in SC-I state and 36.4 GPa in SC-II state, respectively. (c) Upper critical field $H_{c2}$ as a function of critical temperature $T_c$ for 1.7 GPa (purple squares) and 36.4 GPa (orange squares). The $T_c$ at a specific magnetic field is defined from the resistivity criterion of $R_{cri} = 90\%R_n$, where $R_n$ represents the normal-state resistance.

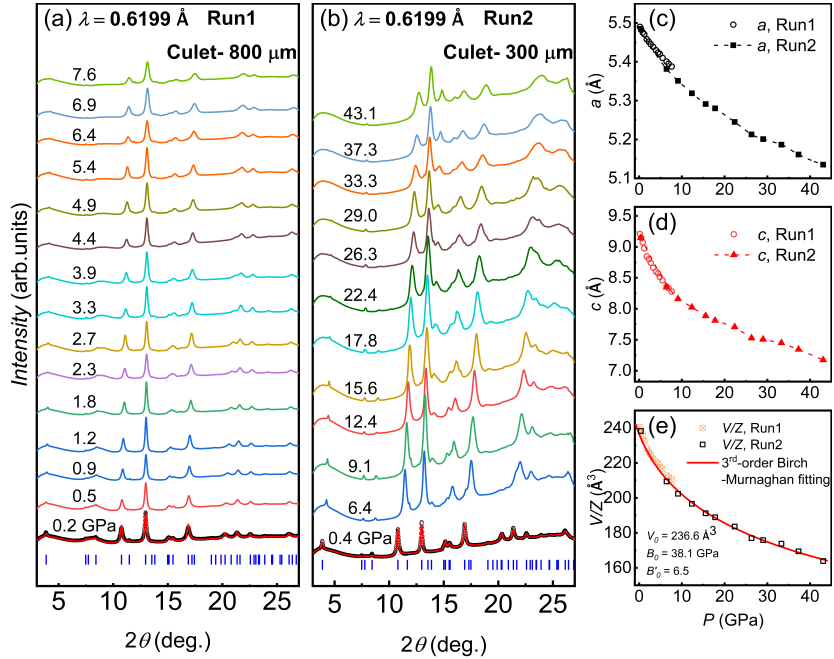

FIG. 4. (a), (b) High-pressure angular dispersive synchrotron XRD patterns of $CsV_3Sb_5$ at pressures ranging from 0.2 to 7.6 GPa (Run1) and from 0.4 to 43.1 GPa (Run2) at room temperature ($\lambda$ = 0.6199 Å) obtained using a DAC with culets of 800 μm and 300 μm in diameter, respectively. The blue tick marks represent the Bragg positions of the hexagonal *P6/mmm* phase of $CsV_3Sb_5$ under the selected fitting results (red lines) at 0.2 (Run1) and 0.4 GPa (Run2). All patterns at various pressures indicate the *P6/mmm* space group for $CsV_3Sb_5$ stabilizing to 47.9 GPa investigated. (c)- (e) Pressure dependence of lattice parameters *a*, *c*, and *V/Z*. The red solid line is the third-order Birch-Murnaghan fitting line.

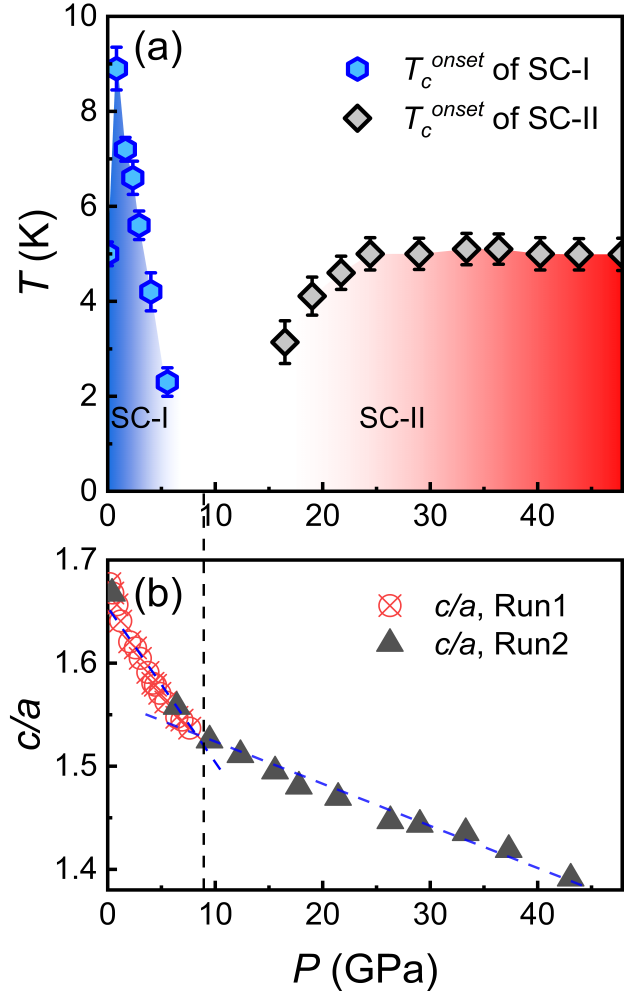

FIG. 5. (a) *T-P* phase diagram of $CsV_3Sb_5$. The onset $T_c$s represented by solid blue hexagons (for SC-I) and solid grey squares (for SC-II). (b) *c/a* ratio as a function of pressure. Red open circles represent data obtained from a DAC with a culet of 800 μm in diameter (Run1), black solid triangles represent data from the 300 μm one (Run2).